\documentclass[showpacs,amsmath,superscriptaddress,reprint,aps]{revtex4-1}

\usepackage{graphicx}
\usepackage{hyperref}

\begin{document}

\title
{Hexagonal boron nitride as an ideal substrate for carbon nanotube photonics}

\author{N.~Fang}
\affiliation{Nanoscale Quantum Photonics Laboratory, RIKEN Cluster for Pioneering Research, Saitama 351-0198, Japan}
\affiliation{Quantum Optoelectronics Research Team, RIKEN Center for Advanced Photonics, Saitama 351-0198, Japan}
\author{K.~Otsuka}
\affiliation{Nanoscale Quantum Photonics Laboratory, RIKEN Cluster for Pioneering Research, Saitama 351-0198, Japan}
\author{A.~Ishii}
\affiliation{Nanoscale Quantum Photonics Laboratory, RIKEN Cluster for Pioneering Research, Saitama 351-0198, Japan}
\affiliation{Quantum Optoelectronics Research Team, RIKEN Center for Advanced Photonics, Saitama 351-0198, Japan}
\author{T.~Taniguchi}
\affiliation{National Institute for Materials Science, Ibaraki 305-0044, Japan}
\author{K.~Watanabe}
\affiliation{National Institute for Materials Science, Ibaraki 305-0044, Japan}
\author{K.~Nagashio}
\affiliation{Department of Materials Engineering, The University of Tokyo, Tokyo 113-8656, Japan}
\author{Y.~K.~Kato}
\affiliation{Nanoscale Quantum Photonics Laboratory, RIKEN Cluster for Pioneering Research, Saitama 351-0198, Japan}
\affiliation{Quantum Optoelectronics Research Team, RIKEN Center for Advanced Photonics, Saitama 351-0198, Japan}
\email[Corresponding author. ]{yuichiro.kato@riken.jp}

\begin{abstract}
Hexagonal boron nitride is widely used as a substrate for two-dimensional materials in both electronic and photonic devices. Here, we demonstrate that two-dimensional hexagonal boron nitride is also an ideal substrate for one-dimensional single-walled carbon nanotubes. Nanotubes directly attached to hexagonal boron nitride show bright photoluminescence with narrow linewidth at room temperature, comparable to air-suspended nanotubes. Using photoluminescence excitation spectroscopy, we unambiguously assign the chiralities of nanotubes on boron nitride by tracking individual tubes before and after contact with boron nitride. Although hexagonal boron nitride has a low dielectric constant and is attached to only one side of the nanotubes, we observe that optical transition energies are redshifted as much as $\sim$50~meV from the air-suspended nanotubes. We also perform statistical measurements on more than 400 tubes, and the redshifts are found to be dependent on tube diameter. This work opens up new possibilities for all-solid-state carbon nanotube photonic devices by utilizing hexagonal boron nitride substrates.
\end{abstract}

\maketitle

Hexagonal boron nitride ($h$-BN), a two-dimensional (2D) material, has played a key role in the development of van der Waals devices. Because of the atomically thin nature, both electrical and optical properties of 2D materials are extremely sensitive to the substrate they contact \cite{Raja:2019,Dean:2010,Uwanno:2018,Fang:2019}. Compared with conventional three-dimensional (3D)    substrates such as SiO$_2$/Si, $h$-BN is atomically flat with low defect density, thereby enabling substantial improvements of device performance and discovery of novel physical phenomena. In electronic devices, the use of $h$-BN has led to mobility enhancement of single layer graphene \cite{Dean:2010,Purdie:2018} as well as appearance of  superconductivity for double layer graphene at the magic angle \cite{Cao:2018}. Encapsulation by $h$-BN reduces the linewidths of photoluminescence (PL) from transition metal dichalcogenides \cite{Ajayi:2017,Buscema:2014}, and interlayer excitons \cite{Calman:2018,Ciarrocchi:2018} in $h$-BN/WSe$_2$/MoS$_2$/$h$-BN heterostructures are utilized in novel excitonic devices \cite{Unuchek:2018}.

It is natural to envision a continued success of $h$-BN substrates in hybrid-dimensional systems, where heterostructures consisting of materials with mixed dimensions are employed \cite{Jariwala:2016}. In particular, we expect that the advantages of $h$-BN will be most evident in structures involving one-dimensional (1D) carbon nanotubes (CNTs), which would comprise 1D/2D heterostructures. The van der Waals interaction should be similar to graphene/$h$-BN heterostructures, leading to a high-quality atomically flat interface. In comparison to conventional solid-state 3D substrates that cause strong PL quenching of CNTs \cite{Kiowski:2006,Schweiger:2015,Sakashita:2010,Lefebvre:2003}, we anticipate improved optical properties on $h$-BN because of the low defect density and the absence of dangling bonds on the surface. In fact, suppression of spectral fluctuations at low temperatures has been observed \cite{Noe:2018}, suggesting the potential benefits of using $h$-BN as a substrate for CNTs. Making use of $h$-BN effects on excitons in nanotubes and eliminating the strong substrate quenching would open up a pathway to all-solid-state CNT photonics.

Here we investigate two different types of heterostructures consisting of $h$-BN and CNT by using PL excitation (PLE) spectroscopy. In samples where $h$-BN flakes are transferred on individual air-suspended CNTs, the chiralities are unambiguously assigned and the same tubes are tracked for $h$-BN effects. Bright luminescence with a narrow linewidth is observed from the CNTs directly attached to $h$-BN at room temperature, indicating weak substrate quenching and small broadening effect from $h$-BN. Excitonic energies are found to exhibit considerable redshifts as well. We also study CNTs transferred on $h$-BN flakes to obtain statistical information on the energy shifts from over 400 tubes. The redshifts are found to be dependent on the CNT diameter, and smaller-diameter tubes show unexpectedly large $E_{11}$ and $E_{22}$ redshifts of $\sim$50~meV. The observed  magnitude of the shifts are too large to explain by dielectric constants alone, considering that $h$-BN is only attached to one side of the CNT and the other side remains unscreened. The results demonstrate the ideal properties of $h$-BN as a substrate for CNT photonic devices.

\begin{figure}
\includegraphics{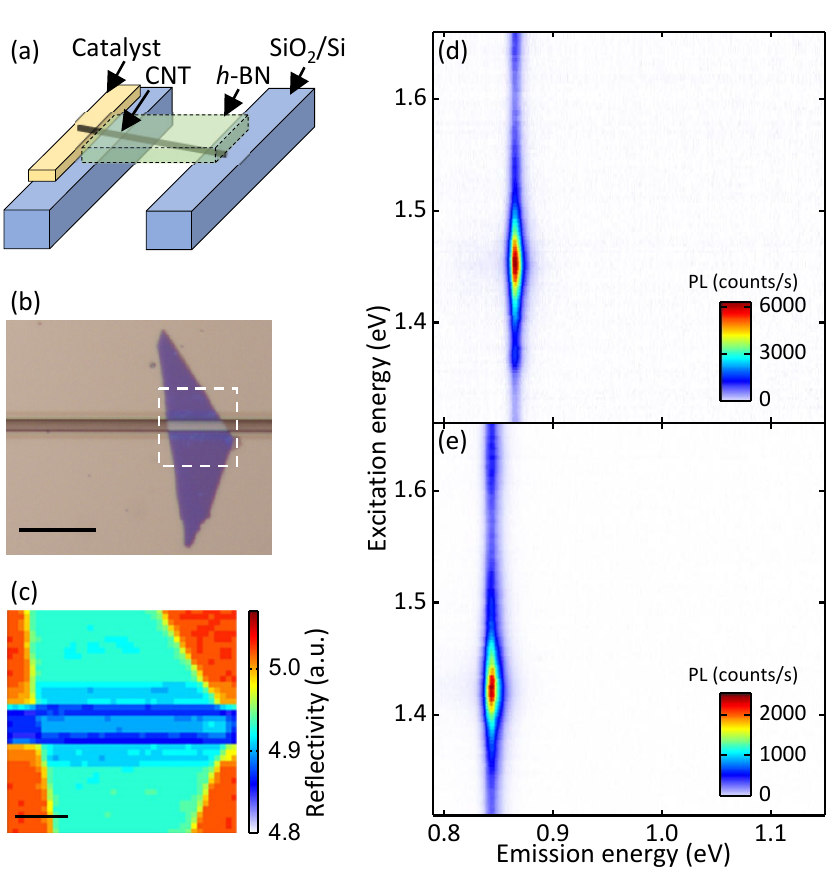}
\caption{
\label{Fig1} (a) A device schematic of an air-suspended CNT with an $h$-BN flake transferred on top. (b) An optical microscope image of a typical $h$-BN/CNT sample. The broken-line box indicates the area measured in (c). (c) A reflectivity image of the $h$-BN/CNT sample shown in (b). Scale bars in (b) and (c) are 20 and 5~$\mu$m, respectively. (d) and (e) PLE maps of an air-suspended (10,8) CNT before and after the transfer of a $\sim$20-nm-thick $h$-BN flake, respectively. The PLE maps are taken with $P=10$~$\mu$W. The linewidths for $E_{11}$ emission before and after the transfer are 9.9 and 12.1~meV, respectively.}
\end{figure}

We begin by studying $h$-BN/CNT heterostructures consisting of an $h$-BN flake on top of an individual air-suspended CNT as shown in Fig.~\ref{Fig1}(a). First, the air-suspended CNTs are prepared on silicon substrates \cite{Ishii:2015}. Electron-beam lithography and dry etching processes are used to fabricate trenches and alignment marks, followed by thermal oxidation to form 70-nm-thick SiO$_2$. Catalyst regions near the trenches are patterned by a second electron beam lithography step, and a 1.5-\AA~Fe film is deposited as a catalyst for the CNT growth by an electron-beam evaporator. The CNTs are synthesized by alcohol chemical vapor deposition at 800$^{\circ}$C for 1~min \cite{Maruyama:2002}. The Fe film thickness is optimized to control the yield such that most of the synthesized CNTs are isolated. Next, an $h$-BN flake is transferred on top of a CNT with the help of the marks on the substrate. The $h$-BN flakes are prepared on polydimethylsiloxane (PDMS) by mechanical exfoliation, and then transferred on the target CNTs at 120$^{\circ}$C by using a micromanipulator system \cite{Fang:2019}. Transferred $h$-BN thickness ranges within 20--90~nm, as determined by atomic force microscopy~(AFM) and optical contrast \cite{Anzai:2019,Bing:2018,Gorbachev:2011}. A typcial $h$-BN/CNT sample is shown in Fig.~\ref{Fig1}(b).

The $h$-BN/CNT structures are investigated with a home-built confocal PL microscopy system at room temperature \cite{Ishii:2015,Watahiki:2012,Jiang:2015,Ishii:2017,Otsuka:2019,Ishii:2019}. A wavelength-tunable Ti:sapphire laser is used for excitation after controlling its power $P$ by neutral density filters, and the polarization is adjusted by a half-wave plate to be parallel with the CNT being measured. The laser beam is focused on the sample using an objective lens with a numerical aperture of 0.8 and a working distance of 3.4~mm. The reflected beam from the sample is detected by a photodiode to map out an image, which can be used to identify the $h$-BN flakes. PL is collected through the same objective lens and detected using a liquid-nitrogen-cooled 1024-pixel InGaAs diode array attached to a spectrometer. All measurements are performed in dry nitrogen to avoid the formation of water-induced defects. 

Figure~\ref{Fig1}(c) shows a reflectivity image of the $h$-BN flake with the target CNT below, reproducing the shape of the $h$-BN flake observed in the optical microscope image [Fig.~\ref{Fig1}(b)]. Using the reflectivity image, we confirm that the nanotube is underneath the $h$-BN flake, and PL spectroscopy is perfomed. A PL image of a tube before and after the tranfer of $h$-BN is shown in Supplementary Figure S1 with no distinct change, implying a homogeous interface between CNT and $h$-BN. 

We characterize the $h$-BN effects on the CNTs by comparing PLE maps of individual CNTs before and after the transfer of $h$-BN [Fig.~\ref{Fig1}(d,e)]. A clean PLE map can be obtained even after the transfer, since the PL intensity remains sufficiently strong. The reduction ratio of PL intensity after transfer compared to before transfer averages to 0.48 over 8 tubes, while the PL reduction ratio is typically less than 0.01 in the current 3D solid-state substrates such as SiO$_2$/Si, quartz, and metals \cite{Kiowski:2006,Schweiger:2015,Sakashita:2010,Lefebvre:2003}. The modest reduction of PL after the transfer of $h$-BN indicates that the substrate quenching effect is significantly suppressed, and such a property makes $h$-BN a promising substrate for light-emitting CNT devices. We point out that the full width at half maximum for $E_{11}$ emission shows only a slight increase by 2.2~meV. In 2D~WS$_2$(WSe$_2$)/$h$-BN heterostructures, the inhomogeneous linewidth broadening of the ground exciton of WS$_2$(WSe$_2$) is reported to be $\sim$2~meV at low temperatures \cite{Raja:2019}. The small change in the linewidth at room temperature suggest that our 1D/2D heterostructures show comparable interfacial properties to 2D/2D heterostructures.

The PLE maps also show substantial modifications of the excitonic energies. We identify the nanotube chirality from the PLE map before transfer [Fig.~\ref{Fig1}(d)] to be (10,8) by extracting the $E_{11}$ and the $E_{22}$ energies from the emission and the excitation resonances, respectively \cite{Ishii:2015}. After the transfer of $h$-BN, we observe $\Delta E_{11}=21$~meV and $\Delta E_{22}=30$~meV, where $\Delta E_{ii}$ is the excitonic energy reduction and $i$ is the index of the transition. Dielectric screening from $h$-BN is expected to cause some redshift, but the observed shifts are large considering the fact that $h$-BN is only attached to one side of the CNT. The excitonic energy shifts are further investigated in detail by taking advantage of $h$-BN/CNT heterostructures where the redshifts can be tracked for individual nanotubes with known chirality.

\begin{figure}
\includegraphics{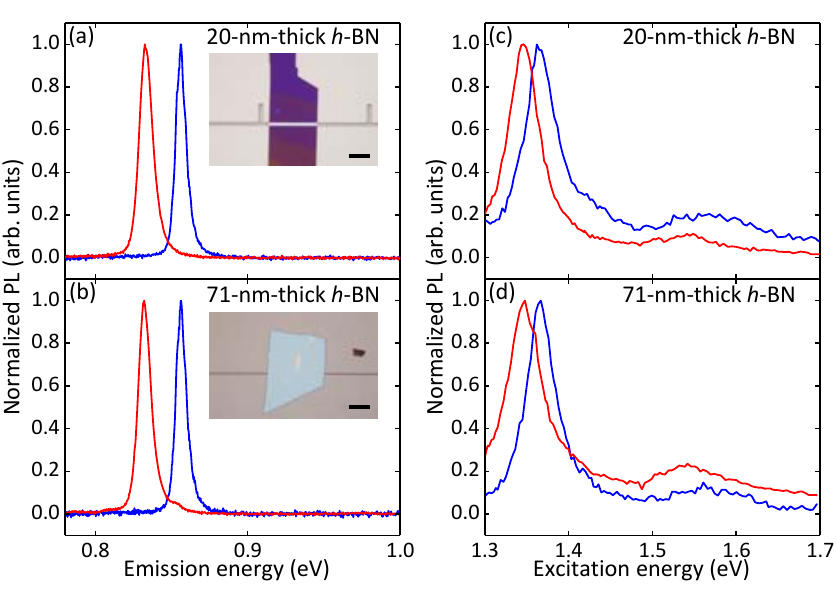}
\caption{
\label{Fig2} (a,b) Integrated and normalized PL spectra for air-suspended (13,5) CNTs before (blue) and after (red) the transfer of $h$-BN flakes with thicknesses of (a) 20~nm and (b) 71~nm. The insets in (a,b) are optical microscope images of the $h$-BN/CNT samples, and scale bars are 20~$\mu$m. (c,d) Integrated and normalized PLE spectra for the same CNTs before (blue) and after (red) the transfer of $h$-BN flakes with thicknesses of (c) 20~nm and (d) 71~nm. $P$ is fixed at 10~$\mu$W.}
\end{figure}

We first consider the sensitivity of the redshifts on $h$-BN thickness. Two $h$-BN/CNT samples are prepared from (13,5) nanotubes using $h$-BN flakes with thicknesses of 20 and 71~nm, which we will refer to as thin and thick $h$-BN samples, respectively [insets of Fig.~\ref{Fig2}(a,b)]. We integrate the PLE maps for each nanotube along the excitation energy axis to obtain PL spectra, and the normalized spectra are shown in Fig.~\ref{Fig2}(a,b). The thin and thick $h$-BN samples show $\Delta E_{11}=23.8$ and $24.4$~meV, respectively. Similarly, PLE spectra shown in Fig.~\ref{Fig2}(c,d) are obtained by integrating the PLE maps along the emission energy axis, where we observe $\Delta E_{22}=18.6$ and $18.7$~meV for the thin and thick $h$-BN samples, respectively. For both $\Delta E_{11}$ and $\Delta E_{22}$, there are no notable changes between the two samples. We have checked the effect of $h$-BN thickness in three more sets of $h$-BN/CNT samples for other chiralities, and the differences in $\Delta E_{11}$ and $\Delta E_{22}$ between samples with thin ($<25$~nm) and thick ($>50$~nm) $h$-BN are found to be less than 3~meV with no clear thickness dependence. The insensitivity of the energy shifts on $h$-BN thickness indicates that 20~nm flakes are thick enough for the full screening of CNT excitons in the perpendicular direction. We note that $E_{11}$ and $E_{22}$ energy shifts can be considered independent of the $h$-BN thickness hereafter, as $h$-BN flakes thinner than 20~nm are not used in this study.

\begin{figure}
\includegraphics{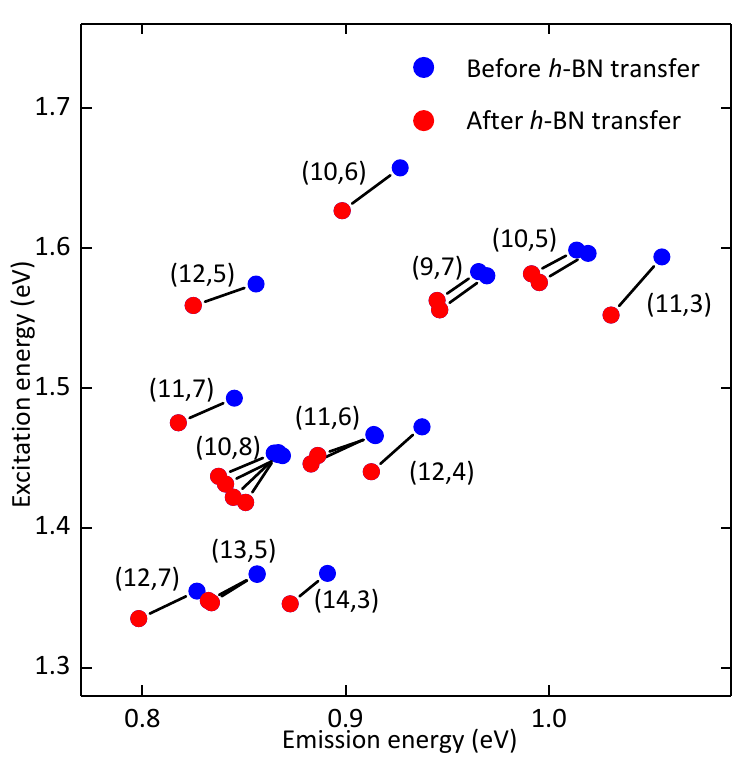}
\caption{
\label{Fig3} PLE peak energies of 19 \textit{h}-BN/CNT samples before $h$-BN transfer (blue dots) and after $h$-BN transfer (red dots) obtained from Lorentzian fits. Lines connect the energies before and after the $h$-BN transfer for each sample.}
\end{figure}

As the redshifts are not sensitive to the $h$-BN thickness, the general behavior of the excitonic energy shifts caused by $h$-BN can be identified by measuring more individual tubes with different chiralities. $E_{11}$ and $E_{22}$ energies before and after the transfer of $h$-BN are extracted and plotted in Fig.~\ref{Fig3}. All 19 CNTs show redshifts in $E_{11}$ and $E_{22}$, which is consistent with the dielectric screening effect. The average $\Delta E_{11}$ and $\Delta E_{22}$ are 24.3 and 22.0~meV, respectively, comparable to the energy difference of micelle-wrapped tubes with respect to air-suspended tubes \cite{Weisman:2003}. Again, the values of the shifts are considered large, as $h$-BN only sits on top of the CNT and the bottom side of the CNT should be unscreened. We also observe some tube-to-tube variations of the shift values within the same chirality. The (10,8) CNTs show an anticorrelated dispersion of the $E_{11}$ and $E_{22}$ energies after the transfer, which may be caused by strain \cite{Yang:1999,Huang:2008,Souza:2005,Koskinen:2010}.

\begin{figure*}
\includegraphics{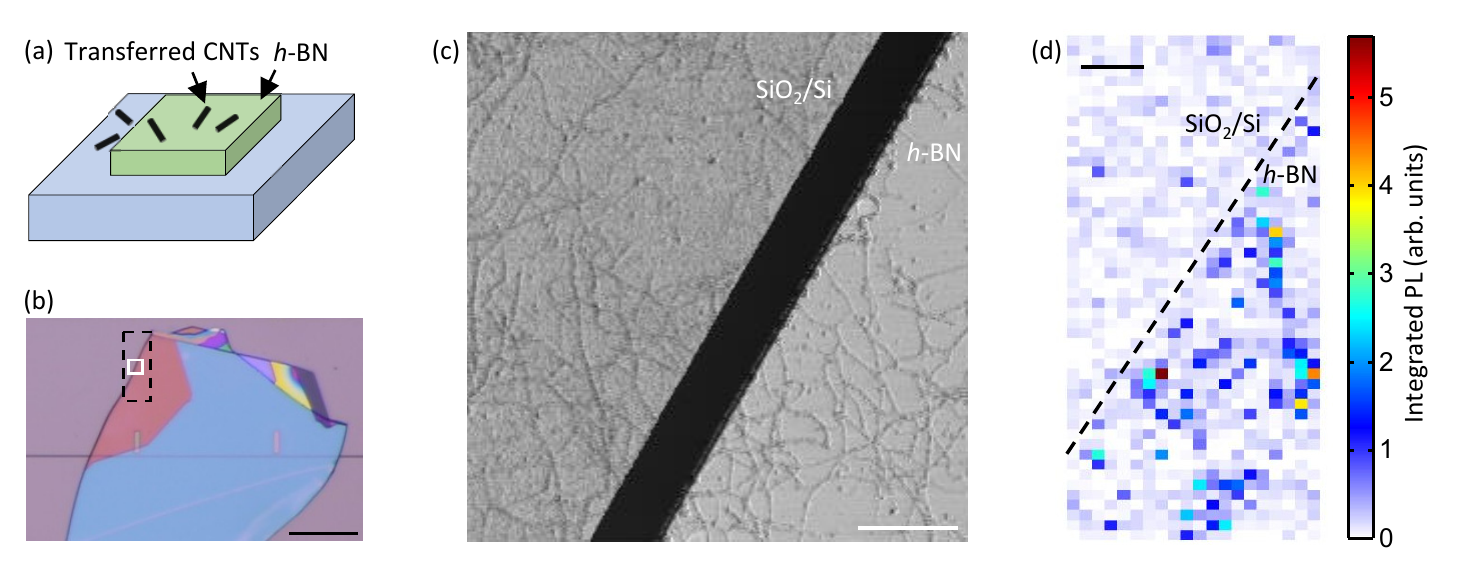}
\caption{
\label{Fig4} (a) A schematic of CNTs transferred on an $h$-BN flake. (b) An optical microscope image of the CNT/$h$-BN sample. White and black boxes indicate the areas measured in (c) and (d), respectively. (c) An AFM phase image at the edge of the $h$-BN flake. (d) An integrated PL image at the edge of the $h$-BN flake. In order to detect nanotubes of various chiralities, four PL images are taken using excitation energies at 1.442, 1.512, 1.570, and 1.699~eV with $P=100$~$\mu$W. The four images are constructed by integrating the emission spectra, and the sum of the PL images is shown here. The boundary of the $h$-BN flake as confirmed by a reflectivity measurement is indicated by a broken line. Scale bars in (b), (c), and (d) are 50, 2, and 4~$\mu$m, respectively.}
\end{figure*}

Up to this point, we have been focusing on $h$-BN/CNT samples with individual nanotubes that allow for chirality identification before $h$-BN transfer. Now that we understand the general behavior of the CNTs upon contact with $h$-BN, we are in a good position to study an ensemble of CNTs directly transferred onto $h$-BN without prior characterization. CNTs are first grown on a SiO$_2$/Si substrate, and we prepare $h$-BN flakes on another SiO$_2$/Si substrate by mechanical exfoliation \cite{Fang:2019}. The CNTs are picked up by using a PDMS/anthracene stamp and transferred on the target $h$-BN flake by using the micromanipulator system. PDMS is peeled off and anthracene is sublimated at 110$^{\circ}$C, leaving a clean surface for CNTs. We will refer to this structure as CNT/$h$-BN, and the corresponding schematic and optical microscope image are shown in Fig.~\ref{Fig4}(a,b), respectively. It is noted that CNTs are transferred on the surrounding SiO$_2$/Si substrate simultaneously due to the large area of the PDMS/anthracene stamp. An AFM phase image of CNTs on both $h$-BN and the SiO$_2$/Si substrate is shown in Fig.~\ref{Fig4}(c), and it is observed that the $h$-BN flake shows smaller surface roughness than that of the SiO$_2$/Si substrate. Many randomly oriented CNTs are firmly attached to both $h$-BN and the SiO$_2$/Si substrate without showing any loose ends or segments. Zoomed-in AFM images are shown in Supplementary Figure S2, where the clean surface of CNTs is observed.

Integrated PL images are taken in the same region as the AFM image and shown in Fig.~\ref{Fig4}(d). Although CNTs are transferred on both the $h$-BN flake and the SiO$_2$/Si substrate, only CNTs on $h$-BN show bright PL. In previous reports of transferred CNTs on quartz or glass substrates, the dominant factor that affects the PL intensity was pointed out not to be the substrate itself but the elimination of interaction between substrates and CNTs by the transfer process \cite{Schweiger:2015}. In contrast, our results reveal that the PL quenching effect is significantly affected by the substrate itself and indicate the superior compatibility between 1D and 2D materials. Compared with 3D substrates, the 2D $h$-BN substrate is atomically flat without dangling bonds. In addition, the van der Waals gap is formed at the 1D/2D interface, which might contribute to the considerable suppression of the substrate quenching effect.

\begin{figure*}
\includegraphics{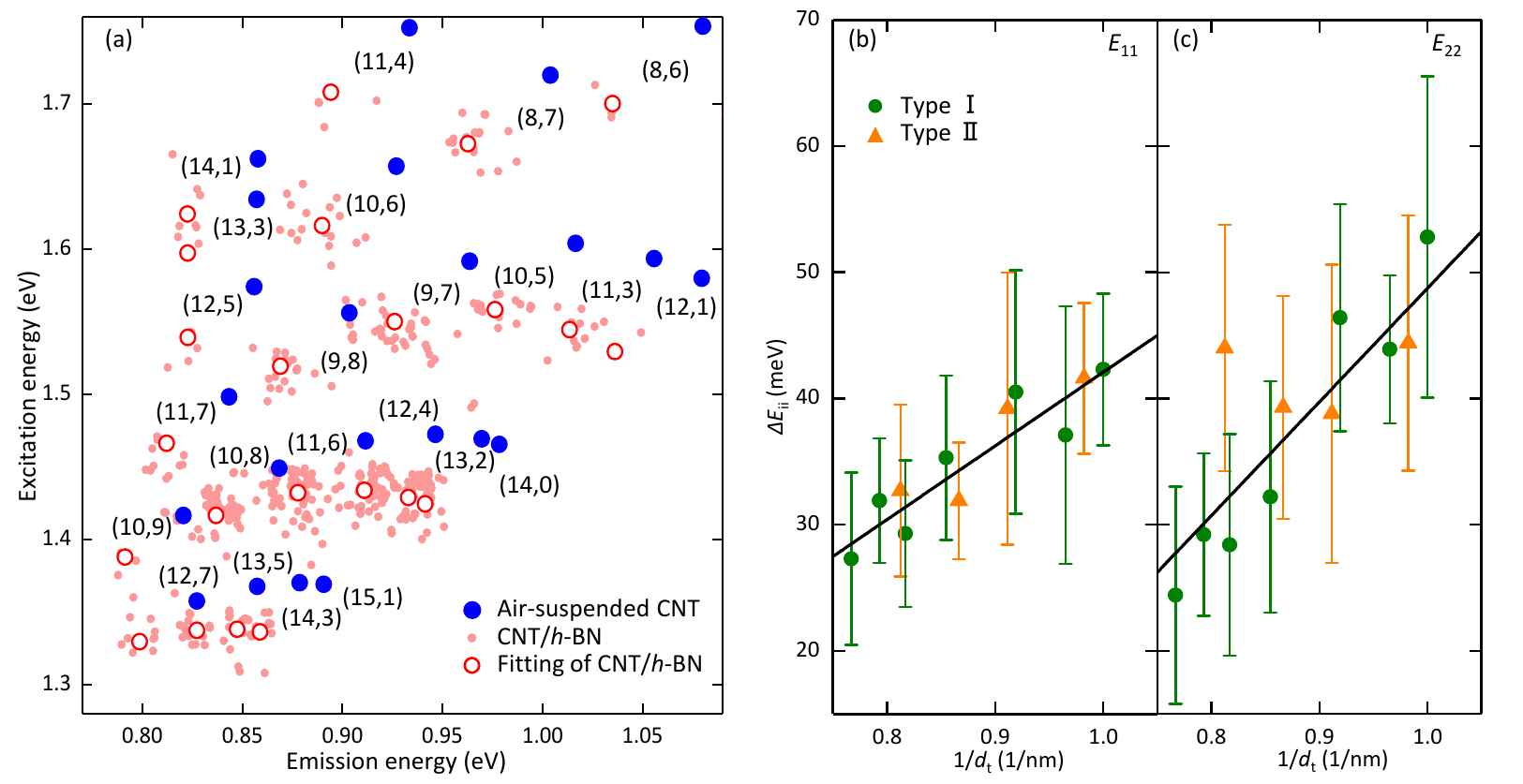}
\caption{
\label{Fig5} (a) PLE peaks of 430 measured CNT/\textit{h}-BN samples (light red dots). Exciton energies for air-suspended CNTs (blue dots) \cite{Ishii:2015} and values for CNT/\textit{h}-BN (red open circle) calculated from Eq.~\ref{equation1} are also shown. (b) and (c) $\Delta E_{11}$ and $\Delta E_{22}$, respectively, as a function of 1/$d_t$ for (8,7), (9,7), (9,8), (10,5), (10,6), (10,8), (11,3), (11,6), (11,7), (12,7), and (13,5) CNTs. The chiralities have been limited to those with unambiguous assignment. Green dots indicate type \uppercase\expandafter{\romannumeral1} tubes while orange triangles indicate type \uppercase\expandafter{\romannumeral2} tubes. Error bars in (b,c) represent standard deviations. }
\end{figure*}

In order to obtain statistical information from the ensemble of CNTs, we have collected more than 400 PLE maps from the CNT/$h$-BN samples. Typical PLE spectra are provided in Supplementary Figure S3. Unlike the CNTs grown over trenches in Fig.~\ref{Fig1}, the transferred CNTs can form bundles because of their random orientation and high density. We observe some PLE maps showing excitation spectra with smeared-out features, and we exclude these PLE maps as they indicate exciton energy transfer in CNT bundles \cite{Tan:2007}. PLE maps with clear features are selected and the $E_{11}$ and $E_{22}$ peak positions are plotted in Fig.~\ref{Fig5}(a). Since the dielectric environment for CNTs in the CNT/$h$-BN structure should in principle be the same as those in the $h$-BN/CNT samples, redshifts in $E_{11}$ and $E_{22}$ are expected to be similar. By examining the correlations with the peak positions in Fig.~\ref{Fig3}, we assign the chiralities for the CNT/$h$-BN sample. In comparison with the $h$-BN/CNT structures, however, the CNT/$h$-BN sample shows slightly larger redshifts for CNTs with the same chirality. The variations of the redshifts are also larger, with a standard deviation of $E_{11}$ in the range of 4--11~meV. The increased redshifts could be attributed to better adhesion between the transferred CNTs and $h$-BN, while the variations might be caused by the formation of bundles in the randomly oriented CNTs as seen in Fig.~\ref{Fig4}(c). The average $E_{11}$ and $E_{22}$ for assignable chiralities are summarized in Table \ref{Table1}.

\begin{table}
\caption{\label{Table1}
Average $E_{11}$ and $E_{22}$ peak energies for carbon nanotubes on $h$-BN obtained from Lorentzian fits of PLE maps. The error values are standard deviations. Some chiralities are not listed, since the large energy variations prevent us from differentiating the chiralities with similar energies such as (13,2) and (14,0).
}
\begin{tabular}{c r@{}c@{}l r@{}c@{}l r@{}c@{}l r@{}c@{}l}
\hline\hline
 & \multicolumn{6}{c}{$E_{11}$} & \multicolumn{6}{c}{$E_{22}$}\\
$(n,m)$ & \multicolumn{3}{c}{(nm)} & \multicolumn{3}{c}{(meV)} & \multicolumn{3}{c}{(nm)} & \multicolumn{3}{c}{(meV)}\\
\hline
(8,6)&	1201.8&	$\pm$&	5.5&	1031.6&	$\pm$&4.7&	729.7&	$\pm$&	5.2&	1699.0&	$\pm$&	12.2\\
(8,7)&	1288.7&	$\pm$&	8.0&	962.1&	$\pm$&6.0&	740.1&	$\pm$&	4.5&	1675.2&	$\pm$&	10.1\\
(9,7)&	1343.7&	$\pm$&	14.1&	922.7&	$\pm$&9.7&	802.5&	$\pm$&	4.7&	1544.9&	$\pm$&	9.0\\
(9,8)&	1424.3&	$\pm$&	8.1&	870.5&	$\pm$&4.9&	818.3&	$\pm$&	5.4&	1515.1&	$\pm$&	10.1\\
(10,5)&	1266.4&	$\pm$&	13.2&	979.0&	$\pm$&10.2&	794.8&	$\pm$&	3.0&	1559.8&	$\pm$&	5.9\\
(10,6)&	1397.1&	$\pm$&	17.0&	887.4&	$\pm$&10.8&	766.3&	$\pm$&	5.6&	1618.0&	$\pm$&	11.8\\
(10,8)&	1478.0&	$\pm$&	10.2&	838.8&	$\pm$&5.8&	873.0&	$\pm$&	5.4&	1420.2&	$\pm$&	8.8\\
(11,3)&	1223.7&	$\pm$&	7.3&	1013.2&	$\pm$&6.0&	804.9&	$\pm$&	6.6&	1540.3&	$\pm$&	12.7\\
(11,6)&	1415.5&	$\pm$&	10.6&	875.9&	$\pm$&6.5&	863.8&	$\pm$&	5.5&	1435.3&	$\pm$&	9.2\\
(11,7)&	1529.7&	$\pm$&	12.8&	810.5&	$\pm$&6.8&	852.8&	$\pm$&	5.7&	1453.8&	$\pm$&	9.7\\
(12,7)&	1550.1&	$\pm$&	13.3&	799.8&	$\pm$&6.9&	930.3&	$\pm$&	6.0&	1332.7&	$\pm$&	8.6\\
(13,5)&	1502.3&	$\pm$&	9.0&	825.3&	$\pm$&4.9&	926.7&	$\pm$&	4.5&	1337.9&	$\pm$&	6.5\\
\hline\hline
\end{tabular}
\end{table}

Statistical results of the $E_{11}$ and $E_{22}$ peak positions in Fig.~\ref{Fig5}(a) allow us to identify other features of the excitonic energy shifts. We find that the energy shifts $\Delta E_{ii}$ compared to air-suspended CNTs \cite{Ishii:2015} is largely affected by CNT diameter $d_t$. Kataura plots of $\Delta E_{11}$ and $\Delta E_{22}$ are shown in Fig.~\ref{Fig5}(b,c) for CNTs with different chiralities, and both $\Delta E_{11}$ and $\Delta E_{22}$ clearly increase with 1/$d_t$. We point out that smaller-diameter CNTs show anomalously large redshifts in $E_{11}$ and $E_{22}$ of $\sim$50~meV. A similar $d_t$ dependence of $\Delta E_{11}$ has been observed for CNTs immersed in organic solvents, which is explained by a model including both external screening from the dielectric environment and internal screening from the CNT itself \cite{Miyauchi:2007}. The screening from the nanotube is weaker compared to $h$-BN, and therefore CNTs with larger $d_t$ will have weaker screening effect and smaller $\Delta E_{11}$. In Fig.~\ref{Fig5}(b,c), we have observed that $\Delta E_{ii}$ is nearly linear with 1/$d_t$, and therefore we describe the data by an empirical equation
\begin{equation}
\label{equation1}
\Delta E_{ii} = A+B/d_t,
\end{equation}
where $A$ and $B$ are fitting parameters. For $\Delta E_{11}$, the values of $A$ and $B$ are $-16.3$~meV and 58.4~meV$\cdot$nm, respectively. For $\Delta E_{22}$, we find $A=-41.4$~meV and $B=90.1$~meV$\cdot$nm. The slope $B$ for $\Delta E_{22}$ is larger than that for $\Delta E_{11}$, implying that excitons from the higher subband are more sensitive to the environment. The electric fields of $E_{22}$ excitons extend further out compared to $E_{11}$ excitons, which might contribute to the higher sensitivity \cite{Araujo:2009}. By using Eq.~\ref{equation1}, we can also deduce $E_{ii}$ peaks of other chiralities that have not been used in the linear fit. The calculated $E_{11}$ and $E_{22}$ peaks based on Eq.~\ref{equation1} are plotted in Fig.~\ref{Fig5}(a) as open red circles, and are located almost at the center of the scattered experimental points for each chirality. 

$\Delta E_{ii}$ is reported to be also dependent on the chiral angles of CNTs. Ohno \textit{et al.} have observed chiral angle dependence of $\Delta E_{ii}$ caused by changes in the effective mass \cite{Ohno:2006prb}, while Jiang \textit{et al.} have calculated that $\Delta E_{ii}$ also has family dependence due to trigonal warping and curvature effects \cite{Jiang:2007}. We have not observed clear chiral angle or family dependence, which might be due to the large energy variations.

Surprisingly, the redshifts of $E_{11}$ in CNTs on the solid-state $h$-BN substrate are comparable to CNTs in solvents with much stronger dielectric screening such as acetone \cite{Ohno:2007}. The in-plane and out-of-plane dielectric constants for bulk $h$-BN are 6.9 and 3.8, respectively \cite{Laturia:2018}, whereas the value for acetone is 20.7. Moreover, CNTs on $h$-BN should not be screened on the top side, while CNTs in acetone are fully surrounded by the solvent. These two factors should lead to reduced screening effects for CNT/$h$-BN in a simple dielectric screening picture. Instead, it appears that the $h$-BN substrate efficiently screens the excitons in the CNTs, and our observation shows that the dielectric constants alone cannot explain the large redshift caused by $h$-BN. It is suggested that there are additional mechanisms contributing to the spectral changes, and further theoretical study would be needed to clarify the origin and to explain Eq.~\ref{equation1}. We note that larger $\Delta E_{ii}$ is expected in 2D materials other than $h$-BN because they usually have larger dielectric constants, which is also promising for the dielectric engineering of CNTs.

In summary, we have investigated $h$-BN effects on the optical properties of CNTs by performing PL spectroscopy on $h$-BN/CNT and CNT/$h$-BN heterostructures at room temperature. We have demonstrated that CNTs directly attached to $h$-BN are highly luminescent with narrow linewidths of  $\sim$12~meV, which is comparable to air-suspended CNTs. The substrate quenching and broadening effects on the 2D $h$-BN substrates are found to be much weaker than those in conventional 3D substrates such as quartz and SiO$_2$/Si. In addition, the anomalously large redshifts in $E_{11}$ and $E_{22}$ of $\sim$50~meV are observed despite the fact that $h$-BN has a low dielectric constant and is only attached to one side of the CNT. The statistical results of PLE maps from over 400 tubes identify the relationship between the redshifts and CNT diameter. These findings highlight the superior properties of $h$-BN for 1D/2D hybrid-dimensional photonics and open a new pathway for manipulating excitons in CNTs.

\section*{Supporting Information}
See supporting information for PL images of an CNT before and after the transfer of $h$-BN, AFM images of CNTs on SiO$_2$/Si and $h$-BN, PLE maps of individual CNTs and bundled CNTs on $h$-BN.

\begin{acknowledgments}
Work supported in part by JSPS (KAKENHI JP19K23593, JP16H05962, JP19H00755), MIC (SCOPE 191503001), and MEXT (Nanotechnology Platform). Growth of hexagonal boron nitride crystals supported by the MEXT Element Strategy Initiative to Form Core Research Center, Grant Number JPMXP0112101001 and JST (CREST JPMJCR15F3). K.~O. is supported by JSPS Research Fellowship. We acknowledge the Advanced Manufacturing Support Team at RIKEN and T.~Nishimura for technical assistance. 
\end{acknowledgments}

\end{document}